# A Survey on the Principles of Persuasion as a Social Engineering Strategy in Phishing

Kalam Khadka
Faculty of Science and Technology
University of Canberra
Canberra, Australia
kalam.khadka@canberra.edu.au

Abu Barkat Ullah
Faculty of Science and Technology
University of Canberra
Canberra, Australia
abu.barkatullah@canberra.edu.au

Wanli Ma
Faculty of Science and Technology
University of Canberra
Canberra, Australia
wanli.ma@canberra.edu.au

Elisa Martinez Marroquin
Faculty of Science and Technology
University of Canberra
Canberra, Australia
elisa.martinez-marroquin@canberra.edu.au

Yibeltal Alem
Faculty of Science and Technology
University of Canberra
Canberra, Australia
yibe.alem@canberra.edu.au

*Abstract*— Research shows that phishing emails often utilize persuasion techniques, such as social proof, liking, consistency, authority, scarcity, and reciprocity to gain trust to obtain sensitive information or maliciously infect devices. The link between principles of persuasion and social engineering attacks, particularly in phishing email attacks, is an important topic in cyber security as they are the common and effective method used by cybercriminals to obtain sensitive information or access computer systems. This survey paper concluded that spear phishing, a targeted form of phishing, has been found to be specifically effective as attackers can tailor their messages to the specific characteristics, interests, and vulnerabilities of their targets. Understanding the uses of the principles of persuasion in spear phishing is key to the effective defence against it and eventually its elimination. This survey paper systematically summarizes and presents the current state of the art in understanding the use of principles of persuasion in phishing. Through a systematic review of the existing literature, this survey paper identifies a significant gap in the understanding of the impact of principles of persuasion as a social engineering strategy in phishing attacks and highlights the need for further research in this area.

*Keywords—Cyber Security, Persuasion Principles, Phishing Email, Social Engineering*

## I. Introduction

Phishing is a form of cybercrime that involves impersonating a trustworthy entity to trick users into disclosing sensitive information or performing malicious actions. This form of cybercrime can be caried out through various channels, such as email, phone, or social media. One of the most sophisticated and effective types of phishing is spear phishing, which targets specific individuals or organizations with personalized and relevant messages. Spear phishing relies on social engineering, which is the manipulation of human psychology to influence decision making.

Moreover, in the social engineering phishing one of the most widely used psychological tricks is applying persuasion principles as social engineering tactics. Persuasion principles are based on the work of Robert Cialdini, who identified six factors that influence human decisions [1]. These principles explain how phishers can manipulate the victims' emotions, beliefs, and behaviors by using various cues and messages. Previous studies have shown that Cialdini's six principles are effective in marketing and sales, additionally also in phishing and other forms of cybercrime [2] [3]. However, there is a lack of consensus on which principles are more prevalent and more impactful in phishing emails, especially in spear phishing. Therefore, further research is needed to explore how persuasion principles are used and perceived in spear phishing emails.

Importantly, a report by [4] indicates that phishing is one of the most common and costly forms of cybercrime, accounting for 22% of data breaches in 2020 [4]. Phishing attacks can target individuals, organizations, and governments, causing financial losses, identity theft, reputation damage, and security breaches. According to the FBI, the number of reported cyberattacks in the US alone reached 791,790 by June 2020, and phishing was the most reported type of cybercrime with 241,342 complaints [5]. The global losses due to business email compromise (BEC) and email account compromise (EAC), which are sophisticated forms of phishing, reached $26 billion between June 2016 and July 2019 [6]. Overall, it has been reported that nearly 1.5 million new phishing websites appear every month, and the most targeted industry sectors are SaaS/webmail, financial institutions, and payment services [7].

In addition, phishing emails have become more personalized and relevant to the recipients, especially in spear phishing [8]. Malicious phishing emails often exploit various principles of persuasion and psychological triggers to influence the victims' decisions [2]. These emails also leverage current events and trends to increase their credibility and appeal, such as the COVID-19 pandemic, cryptocurrency frauds, tax refunds, and online shopping [9, 10].

According to the [11], Australian citizens and businesses lost a record $851 million to scams in 2020, a 23% increase from 2019. Phishing was the most reported scam type in 2020, with 44,084 reports and $1.5 million in losses. Phishing activity increased during the COVID-19 pandemic, especially through government impersonation scams that exploited people's need for information and financial support. In 2021-22, 2.7% of Australians (552,000) experienced a scam, and 8.1% of Australians (1.7 million) experienced card fraud, which was higher than the previous year [12]. In 2019, the Australian Parliament House networks were breached by a cyber-criminal group, suspected to be cyber criminals from a foreign state, in a sophisticated cyberattack that also targeted multiple political parties [13]. In 2020, Service NSW, a government agency that provides online services to citizens, suffered a data breach that exposed the personal information of 186,000 customers and staff [14]. In 2021, Nine Entertainment Co., a media company that owns television channels and newspapers, was hit by a ransomware attack that disrupted its broadcasting and publishing operations [15].

Phishing exploits the gullible nature of human beings and is more successful than other forms of cyber-attacks due to its deceptive nature. Spear phishing has gradually taken over massive spam email distribution and has spread to any possible digital communication format. To illustrate the extent of such exploits, a successful example of a spear phishing attack happened to the Australian National University (ANU) in 2019 [16]. The attack started with the first round of spear phishing against a senior staff member. Through its duration of attacks, which gradually discovered furthermore corporate information, the perpetrators mounted a total of four rounds of spear phishing attacks, with various degrees of success. Eventually, the perpetrators managed to successfully access ANU enterprise systems domain (ESD) of human resources, financial management, student administration and enterprise e-forms systems.

## II. LITERATURE REVIEW

Phishing is one of the main social engineering attacks that aims to exploit human weaknesses and vulnerabilities in the cyber security chain or system processes [17]. According to the definition of [18] phishing is a crime employing both social engineering and technical subterfuge to steal consumers' personal identity data and financial account credentials. Social engineering phishing attacks try to fool unwary victims by making them believe that they are dealing with a legitimate and trusted party using deceptive email addresses, email messages applying different influences and persuasive techniques. It has been shown that phishing emails are created by applying different principles of persuasion as weapons of influence or a strategy of social engineering attacks [2, 19-22]. The principles of persuasion are also called social engineering strategies or weapons of influence. Phishing emails are designed to lead victims to phishing websites that trick the recipient into divulging financial data such as usernames and passwords and other sensitive personal information, which can help in logging into the victims' financial accounts [18]. Therefore, phishing emails are a common attack technique deployed by cybercriminals to lure victims into downloading malware, making them click in malicious links or entering sensitive credentials into a fake login page of the target organisation's own email system [23].

On the other hand, technical subterfuge schemes plant malware onto computers to steal credentials directly, often using systems that intercept victims' accounts usernames and passwords, recording all the activities through keyloggers or web extensions, or alternatively by misdirecting victims to phishing websites [18]. Technical subterfuge is harder than tricking humans now. There have been many studies conducted to explore the technical part of the cyber security, but cybercriminals are seen using persuasive strategy therefore it is clear that the psychological factors need to be considered when studying cyber security measures [24] Social engineers are consistently attempting to use psychological tricks to target humans in an effort to gain unauthorized access to their systems. Humans make decisions based on heuristics; it could be efficient and fast to make a decision, but sometimes there could be decision biases as well and they would make people gullible [25]. As Kevin Mitnick, one of the most famous computer hackers of all time and computer security consultant has mentioned that a company can spend hundreds of thousands of dollars on firewalls, intrusion detection systems and encryption and other security technologies, but if an attacker can call one trusted person within the company, and that person complies, and if the attacker gets in, then all that money spent on technology is essentially wasted and it's essentially meaningless [26].

In addition, social engineering attacks are the psychological manipulation techniques used by cybercriminals to divulge sensitive information from victims by convincing them to act in a way that is not in their interest [27]. Social engineers target victims' human weaknesses or vulnerabilities, using influence techniques aiming at tricking them into clicking a malicious link or disclosing sensitive/confidential information [2] [20] [28]. Due to the advancement of the phishing detection system and other cyber security measures, a human is the weakest link in the cyber security chain as it is hard to get access in the system or get sensitive credentials through the technical breach in the system [29].

## III. METHODOLOGY

The objectives of this survey paper are as follows:

To conduct a comprehensive review of existing research and literature on how the persuasion principles of Cialdini are utilized in phishing studies.

To establish a foundational understanding of the literature that explores the usage of persuasion principles in cybercriminal activities.

To examine the application of popular marketing persuasion principles in the field of cybersecurity, as evident in recent literature.

Search Methods:

The search for relevant articles was conducted using Google Scholar, Scopus, and UC library databases using *persuasion principles of Cialdini in Cybersecurity* keywords. The titles, abstracts, and full-text articles were screened by the researcher. The majority of the records were identified through the initial database searches. Additionally, relevant references were examined, and the researcher's own knowledge and understanding were utilized to identify additional records. Duplicate records were removed, resulting in 50 full-text articles that were assessed for eligibility. Out of these, five articles were excluded as they were not related to phishing literature.

Inclusion Criteria:

To maintain focus on the persuasion principles of Cialdini in cybersecurity, particularly in phishing, the following inclusion criteria were applied. The paper had to be a research study that explored the application of Cialdini's persuasion principles in the context of cyber security. The search was limited to peer-reviewed journal articles and conference presentations written in English. Both quantitative and qualitative studies were considered acceptable.

Exclusion Criteria:

A significant number of articles were initially extracted and screened, but they were subsequently rejected as they did not meet the inclusion criteria. Many of these articles included the term "persuasion" but did not discuss Cialdini's persuasion principles specifically. Additionally, research papers focusing on the other six principles of persuasion within the realm of psychology, rather than cyber security, were also excluded.

## IV. RESULTS AND DISCUSSION

Psychological triggers and principles of persuasion play a significant role in phishing attacks. Many studies have explored the use of psychological triggers, social influence techniques, or principles of persuasion in the context of social engineering and cyber security, with varying results. Most of the work is performed based on the taxonomy proposed by Cialdini [1]. Taib et. al. [30] found that social proof was the most effective strategy, while other studies [21] [31] showed that authority, consistency, and reciprocity were most successful, and social proof and scarcity were the least effective strategies. Cybercriminals target the human element in cyber security, as they are often considered the weakest link. Understanding these principles of persuasion can aid in detecting social engineering attacks, which can be performed through human-based or automated program-based methods [32]. If employees in an organization are aware of these principles and how they are used in phishing attacks, they will be better equipped to detect them. The principles of persuasion consist of six basic tendencies of human behaviour that can generate a positive response [1]. These six principles of persuasion, also known as weapons of influence, are used in various aspects of life, including business dealings, societal interactions, and personal relationships [1].

After reviewing the work of [1, 2, 21, 33, 34] the six principles of persuasion in phishing can be understood as follows:

1. Reciprocation: Reciprocation is the custom or norm that compels individuals to repay in kind what they have received. In other words, to return the favour or adjust to smaller request [1]. This occurs when a phisher sends an email containing a message that is perceived as a request or obligation towards the recipient to "return the favour" [2]. People tend to return a favour, so offering something for free or doing a small favour for someone can lead them to feel obligated to return that favour. For example, if someone offers you a free sample of a product, you may feel more inclined to purchase that product.

2. Commitment/Consistency: Users will honour commitments they have previously made and be consistent with their actions [35]. If a company reminds a user of the terms of use and their agreement to change their password yearly, they may feel committed to these terms and feel obligated to follow the request [2]. For example, if someone has already agreed to a small request, they may be more likely to agree to a larger request or if someone is running a campaign about global warming then they could be tricked into clicking global warming related phishing links easily.

3. Social proof: People will let their guard down and start aligning themselves with society or group behaviour if they believe everyone around them shares the same risk [35]. They also want to be included in what other people are doing. The trick with social proof could include an email referring to other customers who have the same needs and wants as the person who viewed the item with a URL leading them to the malicious page [2] [33]. People often look to the actions of others to determine their own behaviour, so providing examples of others who have taken a certain action can be an effective way to influence behaviour. For example, if we see a crowd of people doing something, we may be more likely to do it as well.

4. Authority: Individuals tend not to question authority. This could be out of fear or to avoid negative consequences such as losing privileges, humiliation, or condemnation. Authority could be displayed as an official signature/logo or an email coming from an administrator [2] [1]. People tend to follow the lead of those they perceive as experts or authority figures, so emphasizing phishers' credibility or expertise can be an effective persuasion technique, or social engineering strategy, in phishing emails.

5. Similarity/Liking: Individuals tend to be easily persuaded by those people whom they know and like, or people who are similar to themselves [1]. Additionally, people trust those they find attractive or credible, and trust increases compliance [35]. If an email is like previous emails from an organization or appears credible, users could trust the message and comply with requests [2] [1] [33] [35]. People are more likely to be influenced by those they admired like celebrity figure so building rapport and highlighting similarities can increase the likelihood of tricking them in a phishing scam.

6. Scarcity: An emotional response is elicited when the availability of an item/service is limited or there is only a short timeframe to respond. Cybercriminals could include a threatening message such as: respond to this email within 12 hours or you lose access to your email account, you will have money deducted from your bank account, or a multitude of other consequences [2] [1] [33] [35].

*A. The study on principles of persuasion and phishing*

Phishing susceptibility has been the subject of numerous studies [23, 36-40], with the focus being on identifying psychological triggers and persuasion strategies that make individuals vulnerable to these types of cyber-attacks. Wang et al. [41] conducted a phishing susceptibility study that was based on the theoretical structure of the Theory of Deception by [42]. This further exploration of phishing revealed that attention to visceral triggers, such as emphasizing the urgency to respond, increases the likelihood of an individual responding to a phishing email, while attention to phishing deception indicators, such as grammar errors or the sender's address, decreases the likelihood of a response. Furthermore, the study found that the cognitive effort expended in processing a phishing email is not significantly related to the likelihood of a response, after controlling for both attentions [41].

Ferreira and Teles [21] integrated principles of persuasion for social engineering attacks by drawing on popular principles of persuasion from marketing and scams. The researchers built on the foundational work of Cialdini's, Gragg's, and Stajano & Wilson's to derive a unique list of principles of persuasion in social engineering attacks. The final list of principles includes Authority (Auth), Social Proof (SP), Liking, Similarity & Deception (LSD), Distraction (DIS), and Commitment, Reciprocation & Consistency (CRC). These principles demonstrate how phishers use psychological triggers and persuasion strategies to manipulate individuals into giving away sensitive information or taking specific actions. For example, an email claiming to be from a recipient's bank, including the bank name in the subject line, is an example of how phishers use the principle of Authority.

Consequently, the principles of persuasion in phishing have been studied through different methods in cyber security fields. Early publications are more focused on the content analysis of phishing emails to find out the behaviour of adversaries, and which persuasion principle is more effective in tricking through phishing emails. Also, content analysis of differently themed phishing scenarios has been done based on the principles of persuasion e.g., content analysis of the coronavirus-themed phishing emails [10], cryptocurrency fraud cases [9] and vishing scenarios [28]. Later, it has been studied to understand the susceptibility of phishing emails in the university environment, public administration environment and organizational settings. Additionally, there are studies which are focused on providing training, education and awareness to users and the effectiveness of these training materials when using the principles of persuasion as an attack vector. In those studies, the effectiveness of gamification of training materials, embedded training materials, simulations, role-playing methods and cyber security awareness training have been conducted. Moreover, the principles of persuasion were studied to implement a phishing detection method based on these principles using machine learning and Artificial Intelligence (AI) techniques [43].

*B. Phishing content analysis*

Many of the available detection models depend on email descriptions like domain names, URLs, and associated email addresses. Whilst these detection models and techniques which rely on email metadata are gradually reducing and helping to mitigate phishing scams, many phishing attacks continue to manipulate victims by misusing human vulnerabilities. Therefore, the researchers have studied the content of these phishing emails to understand how cybercriminals are manipulating human vulnerabilities and using persuasion tactics to encourage their targets to divulge sensitive information and personal credentials.

The study performed by [2] is the first research conducted to understand adversaries' behaviour patterns by considering the six principles of persuasion by Cialdini, how phishing emails are constructed using these principles of persuasion and which persuasion principle is most widely used in phishing emails. In addition, [44] conducted a similar study by applying the principles of persuasion in the phishing email subject line. It has been recommended that a phishing detection system, considering the email subject line, could be very effective in detecting and preventing the phishing emails.

Olga A Zielinska et. al. [35] conducted a temporal analysis of the principles of persuasion in phishing emails and they further explored correlational analysis of email characteristics by year, and it revealed that the principles of persuasion of commitment and scarcity have increased over time.

Similarly, Jones et. al. [28] studied how social engineers use the principles of persuasion during vishing attacks. This study has found that authority, social proof, and distraction were the most used principles in vishing types of social engineering attacks followed by liking, similarity, and deception. Weber et al. [9] conducted content analysis of cryptocurrency fraud cases using the principles of persuasion. They tried to find out how the principles of persuasion are used to exploit cryptocurrency users. This study analysed which psychological tricks had been used by the phishers in cryptocurrency fraud cases.

The content analysis of Covid-19 themed phishing emails was performed in [10]. This empirical study explored the influence methods, fear appeals, and urgency cues applied by phishers to trick or coerce users to follow instructions presented in coronavirus-themed phishing emails. In this study 208 coronavirus-themed phishing emails were examined, and content analysis was performed based on fear appeals, urgency cues, source credibility and Cialdini's principles of persuasion. This content analysis of the coronavirus-themed phishing emails revealed that authority, commitment, liking, fear appeals, and urgency cues were the most frequently employed influence methods.

*C. Phishing susceptibility study based on principles of persuasion*

This section reviews the publications which are focused on understanding the effect of psychological factors on the susceptibility of phishing targets. For example, [19] examined the influence of three social engineering strategies on users' judgments of how safe it is to click on a link in an email. The three strategies examined were authority, scarcity and social proof, and the emails were either genuine, phishing or Spear phishing. Of the three strategies, the use of authority was the most effective strategy in convincing users that a link in an email was safe. When detecting phishing and Spear phishing emails, users performed the worst when the emails used the authority principle and performed best when social proof was present. Overall, users struggled to distinguish between genuine and Spear phishing emails. Finally, users who were less impulsive in making decisions were generally less likely to judge a link as safe in the fraudulent emails.

Additionally, [22] conducted a study to investigate whether an interaction between personality and the principles of persuasion used in the phishing emails. They applied four of Cialdini's principles of persuasion and confirmed that high extroversion is predictive of increased susceptibility to phishing attacks. Their study also identified the most effective persuasion principle to utilize in both phishing and legitimate emails was "liking". Conversely, the combination of authority & scarcity principles of persuasion was most likely to arouse suspicion in both phishing and legitimate emails.

Oliveira et al. [38] conducted a field experiment investigating spear phishing susceptibility as a function of internet user age (old vs young), weapon of influence, and life domain. This study focused on which weapon of influence and which life domains are more effective, in addition to how the susceptibility varies with respect to the age and gender of targets. This study revealed that older women were the most vulnerable group to phishing attacks, while younger adults were most susceptible to scarcity, as well as the fact that older adults were most susceptible to reciprocation.

Lawson et al. [22] investigated the relationship or interaction between principles of persuasion and the personality of victims. Additionally, the signal detection theory framework was used to evaluate how principles of persuasion are present in phishing email. It has concluded that people with highly extroversion personalities are highly predictive of susceptibility to phishing emails. Furthermore, it has also concluded that there are interaction effects between the personality of the targets and the persuasion principle utilized in the context of phishing emails. The limitation of this study was that the participants were suspicious about phishing study; therefore, phishing detection accuracy has higher than legitimate detection accuracy.

De Bona and Paci [23] investigated which persuasion technique, between authority and urgency, is more effective in making employees susceptible to phishing. They have also studied the relationship between employees' susceptibility and their demographic data, and the effectiveness of embedded training in reducing employees' susceptibility to phishing attacks. It has found that employees were more vulnerable to phishing attacks when the urgency principle was exploited.

Frauenstein and Flowerday [25] examined the effect of the relationship between the Big Five personality model and the heuristic-systematic model of information processing. This study has revealed that conscientious users were found to have a negative influence on heuristic processing and are thus less susceptible to phishing on social network sites (SNSs). The study also confirmed that heuristic processing of decision-making process increases susceptibility to phishing.

Aljeaid et al. [36] studied the level of cyber security knowledge and cyber awareness in Saudi Arabia. It found that the most effective influence techniques used in phishing attacks are commitment/consistency, liking, authority, and scarcity. It also demonstrated that users can easily fall victim to phishing attacks if they are not sufficiently knowledgeable to make good judgments in cyberspace. This study suggested that cyber security awareness should become a new culture and must be taught at a very young age to improve cyber awareness and develop sustainable safe cyber behaviour.

*D. Training, awareness and education of cyber security threats using principles of persuasion*

The publications which focused on training, awareness and education of cyber security threats using persuasion principles:

Sharevski and Jachim [45] have conducted an empirical study investigating the effectiveness of using an intelligent voice assistant, Amazon Alexa. They have developed interaction-based phishing training focused on the principles of persuasion with examples on how to detect them in phishing emails. This study results showed that the participants in the interaction-based group outperformed the others when detecting phishing emails that employed the persuasion principles. The persuasion principles which were employed in this study were: authority, authority/scarcity, commitment, commitment/liking, and scarcity/liking.

De Bona and Paci [23] have suggested that the most common approach to mitigate phishing attacks is employees' education and awareness. Embedded training can be used to make employees' aware about phishing attacks that can educate employees when they fall for the attacks. This study has shown that embedded training was perceived as effective by employees, but it did not reduce their susceptibility to phishing attacks. Additionally, they have found that employees were more vulnerable to phishing attacks when urgency principle was exploited in phishing emails compared to the authority principle.

Pirocca et al. [46] has presented a toolkit to support organizations in offering their employees security awareness training against tailored spear phishing attacks. This study has demonstrated the toolkit by showing how it can be used to address the limitations in the experiment design presented in [38], it has utilised the principles of persuasion and life domain to tailor the spear phishing emails.

Fatima et al. [47] have introduced a systematic approach to design serious games for security education. Their results have shown that awareness of phishing risks is improved and the resistance to potential first attack contact is enhanced. The game showed a positive effect on participants' understanding of excessive online information disclosure.

*E. Machine learning and phishing*

Machine learning is an essential technology that plays a crucial role in detecting phishing attacks, as it can rapidly process and examine large quantities of data to identify possible threats [48]. The publications which are experimented on machine learning techniques by applying the principles of persuasion in phishing are:

Li et al. [43] have described the use of the principle of persuasion in phishing emails and proposed a phishing email detection method based on the persuasion principle using machine learning techniques.

Valecha et al. [49] have created three machine learning models with relevant gain persuasion cues, loss persuasion cues, and combined gain and loss persuasion cues, respectively, and compared them with a baseline model that does not account for persuasion cues. The study shows that using persuasion cues can improve the design of effective countermeasures for detecting and blocking phishing emails.

*F. Latest trends on the study*

The study of the principles of persuasion in phishing emails has evolved from content analysis to assessing susceptibility of targets, with a focus on understanding adversary behaviors through email content and analyzing the responses of targets based on their personalities. The latest trends of publications on the principles of persuasion as a social engineering strategy can be taken example of study conducted in [50].

Xu et al. [50] have designed a synthetic simulation environment to simulate spear phishing and tried to understand the cognitive processes from both the adversarial and end-users' perspectives. It has found that end-users were more vulnerable to spear phishing attacks as more information was exposed to attackers and it has revealed that attackers tended to create narratives and impersonation contextually when accessing more information about their targets.

## V. GAPS IN CURRENT RESEARCH

This literature review shows a few limitations and gaps in the current research on how principles of persuasion are utilised as social engineering strategy in phishing. There is a lack of long-term studies about the susceptibility of victims and the behaviour of adversaries; many studies on phishing emails have focused on studying short-term effects and using email samples from short time frame [2, 21, 30] and have not examined the long-term impact of these phishing emails and pattern of changes of these on organizations and individuals. Long-term studies could provide valuable insights into the ongoing effects of these phishing emails. There is a lack of participation from various demography or groups of people such as different regions, cultures, age, study background, experiences in the study as most of the study have been conducted in the university environment [19, 37, 45, 51], even if some studies [23, 39, 47] are conducted on the organisational environment there is lack of study in other variation of organisational environment for instance in small financial organisation, IT service environment, industrial environment etc. Because of that the results are not generalizable to other regions or cultures or in organization environment. In most research studies, the findings are limited in their generalizability due to the small sample sizes used. Additionally, this is a concern that researchers need to be mindful of when interpreting and applying the results to broader populations or contexts. There is limited understanding for the effects of combination of different principles of persuasion because most of the study has experimented the effectiveness of an individual principles of persuasion but there is a lack of study on how those can be used in combination to increase the effectiveness of phishing attacks. Another limitation and gap in the study of principles of persuasion as a social engineering strategy is the lack of understanding the mindset of victims or targets of phishing attacks after being tricked by cybercriminals using those principles because it can provide valuable insights into the effectiveness of different types of phishing attacks, and how they can be prevented in the future by researching effective strategies to counter phishing and improve overall cyber security. Much of the research on phishing emails have been based on simulated or artificial phishing emails rather than real-world examples. This has limited the generalizability of the findings to real-world scenarios. The previous studies have limited focus on the attackers' motivations, tactics, and characteristics of the attackers. There is also a limited focus on certain types of social engineering attacks: research has mostly focused on phishing emails while other types of attacks such as baiting, spear phishing emails, whaling, and impersonation are not well studied. There are also gaps of the performing systematic evaluation of defence strategies from social engineering attacks. When experimenting, there is also a lack of control groups because of that it is making difficult to determine the effectiveness of the intervention or to compare the results to a non-intervened group. Last but not least, there is also a lack of standardization in the coding of principles of persuasion, measurement methodologies and data because of that, there are different results in different study.

## VI. CONCLUSION

This paper identified a gap in understanding the impact of principles of persuasion in phishing attacks and the need for more research. Phishing emails often use social engineering tactics to manipulate victims into divulging sensitive information and unauthorized access in system, this survey concluded that phishing emails which have applied principles of persuasion are more effective social engineering tactics than other phishing attacks. Spear phishing, a targeted form of phishing, is particularly effective when using these persuasion techniques. In recent years, there has been a growing body of research on the use of principles of persuasion in phishing emails, as well as the relationship between principles of persuasion and phishing susceptibility.